\documentclass[10pt, aps,prl,twocolumn,showpacs,superscriptaddress,amsmath,amssymb]{revtex4-1}

\usepackage{graphicx}
\usepackage{hyperref}

\usepackage{color}
\usepackage[normalem]{ulem}

\graphicspath{{figures/}}

\newcommand{\eq}[1]{Eq.~\eqref{eq:#1}}

\newcommand{\fig}[1]{Fig.~\ref{fig:#1}}

\usepackage{SIunits}
\newcommand{\myunit}[1]{\unit{#1}{\micro\meter}}
\newcommand{\myunitinv}[1]{\unit{#1}{\micro\reciprocal\meter}}

\hypersetup{bookmarksnumbered=true,
  bookmarksopen=true,
  citecolor=blue,
  colorlinks=true,
  hypertexnames=true,
  linkcolor=blue,
  linktocpage=true,
  pdfauthor={Thomas Hisch, Matthias Liertzer, Dionyz Pogany, Florian Mintert, Stefan Rotter},
  pdfhighlight=/I,
  pdfpagemode=UseNone,
  pdftitle={Pump-controlled light emission from random lasers},
  urlcolor=blue}

\definecolor{blau}{rgb}{0.9,0,0.2}

\newcommand{\bmn}{\beta_{m,n}}
\newcommand{\bmnwb}{\{\bmn\}} \newcommand{\xvec}{\mathbf{x}}

\definecolor{darkgreen}{rgb}{0,0.5,0}
\definecolor{grey}{rgb}{0.8,0.8,0.8}

\newcommand{\gT}{\gamma_{\bot}}

\newcommand{\gc}{gain curve}

\newcommand{\ff}{far-field}

\newcommand{\fTLM}{\bar{\Psi}_{1}}

\newcommand{\pp}{pump profile}

\begin{document}

\title{Pump-Controlled Directional Light Emission from Random Lasers}

\author{Thomas Hisch}
\email{t.hisch@gmail.com}
\affiliation{Institute for Theoretical Physics, Vienna University of
             Technology, Wiedner Hauptstra\ss e 8--10/136, 1040 Vienna, Austria, EU}

\author{Matthias Liertzer}
\affiliation{Institute for Theoretical Physics, Vienna University of
             Technology, Wiedner Hauptstra\ss e 8--10/136, 1040 Vienna, Austria, EU}

\author{Dionyz Pogany}
\affiliation{Institute for Solid State Electronics, Vienna University of Technology, Floragasse 7, 1040 Vienna, Austria, EU}

\author{Florian Mintert}
\affiliation{Freiburg Institute for Advanced Studies, Albert--Ludwigs University of Freiburg, Albertstra\ss e 19, 79104 Freiburg, Germany, EU}

\author{Stefan Rotter}
\email{stefan.rotter@tuwien.ac.at}
\affiliation{Institute for Theoretical Physics, Vienna University of
              Technology, Wiedner Hauptstra\ss e 8--10/136, 1040 Vienna, Austria, EU}

\date{\today}

\begin{abstract}
  The angular emission pattern of a random laser is typically very
  irregular and difficult to tune. Here we show by detailed numerical
  calculations that one can overcome the lack of control over this
  emission pattern by actively shaping the spatial pump
  distribution. We demonstrate, in particular, how to obtain
  customized pump profiles to achieve highly directional emission.
  Going beyond the regime of strongly scattering media where localized
  modes with a given directionality can simply be selected by the
  pump, we present an optimization-based approach which shapes
  extended lasing modes in the weakly scattering regime according to
  any predetermined emission pattern.
\end{abstract}

\pacs{42.55.Zz, 42.25.Fx}
\maketitle

The scattering of waves in a disordered medium is a ubiquitous
physical process that occurs in a variety of different scenarios
ranging from the micro- to the macroscale
\cite{akkermans_mesoscopic_2007,letokhov_astrophysical_2008}. Although
much attention has been devoted to this topic, new intriguing
phenomena continue to emerge in this area of physics
\cite{barthelemy_levy_2008,*sapienza_cavity_2010,*wang_transport_2011,*baudouin_cold-atom_2013}.
Particularly exciting progress has recently been reported from the
field of disordered photonics \cite{wiersma_disordered_2013}: randomly
scattering materials with optical gain were shown to operate as lasers
despite the absence of a cavity
\cite{cao_lasing_2003,*wiersma_physics_2008,*zaitsev_recent_2010}.  In
contrast to conventional lasers, however, such {\em random lasers}
emit light at multiple frequencies and with a broad spatial emission
profile
\cite{cao_lasing_2003,*wiersma_physics_2008,*zaitsev_recent_2010,Cao,tureci_strong_2008,*TurStoRot2009}.
On the other hand, unprecedented control of light fields in complex
media has been achieved through optical {\it wave front shaping} based
on spatial light modulators \cite{mosk_controlling_2012}.  This
progress has, e.g., enabled the spatial and temporal focusing of light
or the projection and retrieval of images through strongly scattering
media
\cite{vellekoop_focusing_2007,popoff_image_2010,*popoff_measuring_2010,KatSmaBro2011,AulGjoJoh2011,rotter_generating_2011}.

In the present Letter we strive to bridge the advances in the above
two fields in order to tackle the longstanding goal
\cite{wiersma_light_2001,*lee_laser_2002,*wu_random_2004,*gottardo_quasi-two-dimensional_2004,*vanneste_localized_2005,*savels_gain_2007,*gottardo_resonance-driven_2008,*fujiwara_numerical_2009,*liang_zno_2010,*bardoux_single_2011,*el-dardiry_tuning_2011,*frank_transport_2012,*stano_suppression_2013,leonetti_mode-locking_2011,bachelard_taming_2012,bachelard_turning_2013,leonetti_active_2013}
of making random lasers externally tunable.
\begin{figure}
  \centering
  \includegraphics{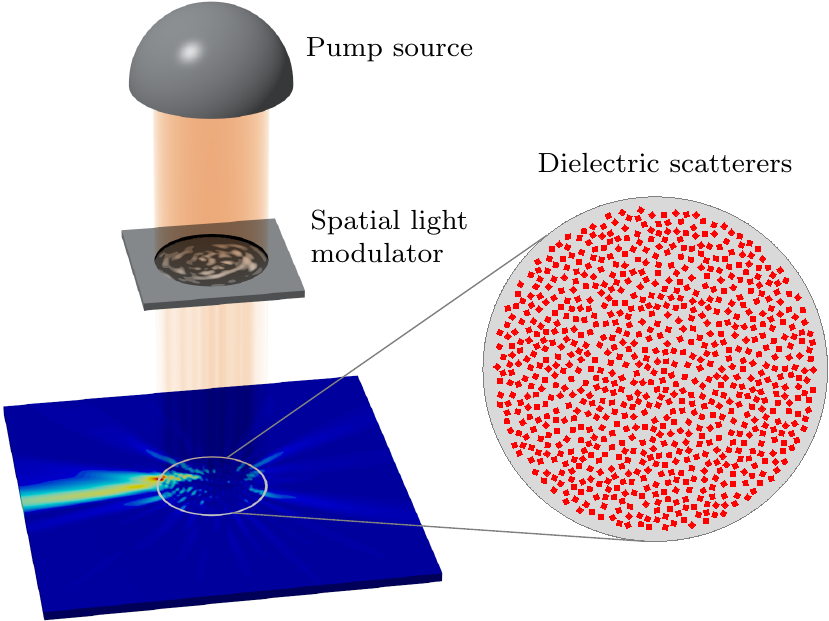}
  \caption{(Color online) Sketch of the envisioned setup for realizing
  pump-controlled emission from a random laser: Before being
  directed to the random medium, an applied pump beam is
  sent through a spatial light modulator (SLM) which shapes the
  beam following an optimization procedure. The targeted emission
  pattern can be chosen at will. The geometry of the disk-shaped
  random laser (radius $R=\myunit{1}$) consists of square shaped scatterers with index $n_s$ (red),
  embedded in a uniform background medium with index $n_b=1$ (gray). 
 }
  \label{fig:geom}
\end{figure}
Employing a quasi-one-dimensional (1D) laser, it was recently shown
numerically \cite{bachelard_taming_2012} as well as experimentally
\cite{bachelard_turning_2013} that, using wave front shaping
techniques to iteratively optimize the pump profile applied to a
random laser, a single-mode lasing operation can be achieved at a
predetermined frequency.  With this approach one of the major
deficiencies of a random laser, i.e., the lack of control over its
emission frequency could be remedied. This leaves random lasers with
the major drawback of emitting into arbitrary directions -- a property
which severely limits their applicability in many practical
circumstances. Ideally, of course, the control over the angular
emission pattern could also be exerted through the applied pump
profile which is the most easily tunable control knob in a random
laser. Demonstrating this explicitly seems, however, a formidable
theoretical challenge. Consider, for comparison, the complexity
involved in previous demonstrations of pump optimization for random
lasers
\cite{leonetti_mode-locking_2011,bachelard_taming_2012,bachelard_turning_2013,leonetti_active_2013}
or in the well-studied problem of engineering a microcavity laser with
unidirectional emission
\cite{GmaCapNar98,*WieHen2008,*YiKimKim2009,*YanWangDiehl2009,*SonGeSto2010,*WanYanYu2010,*shinohara_chaos-assisted_2010}.
In both of these two cases, the functions which were optimized in one
or several sequential steps were strictly one-dimensional: For the
random lasers this function is related to the spatial pump profile and
for the microcavity lasers it is the cavity boundary. In contrast, for
fully controlling the emission pattern of a random laser one has to
deal with an essentially two-dimensional pump profile with a very
large number of adjustable degrees of freedom.

We address this challenge numerically, employing standard
semiclassical laser theory at lasing threshold
\cite{haken_light:_1985}. Our dielectric random laser medium consists
of a passive and active component, $\varepsilon=\varepsilon_c+\chi_g$,
with the passive dielectric function $\varepsilon_c$ given by a static
and randomly generated as well as immobile distribution of
square-shaped sub-wavelength scatterers (see
Fig.~\ref{fig:geom}). These scatterers, which all have
the same index of refraction $n_{s}$, are suspended in a disk-shaped
background material of index $n_{b}=1$. Both the scatterers and the
background medium are assumed to be optically active, with a quantum
susceptibility $\chi_g$ that depends on the wave number $k$
\cite{siegman_lasers_1986},
\begin{equation}
  \chi_g = D_0 F(\mathbf{x}) \gamma(k),\,\,\,\,\, {\rm with}\,\,\, \gamma(k) = \frac{\gamma_\bot}{k - k_a + i \gamma_\bot}.
\end{equation}
This active component is subject to a position-dependent external pump
source $F(\mathbf{x})$ as realized, e.g., with an optical pump beam
controlled by a spatial light modulator (see schematic setup shown in
Fig.~\ref{fig:geom}). Here, $D_0$ defines an overall pump strength,
$k_a$ is the frequency of the atomic transition (we consider the speed
of light $c=1)$ and $2\gamma_\bot$ is the width of the
Lorentzian-shaped gain curve $|\gamma(k)|^2$. Note, that this approach
is identical to the noninteracting form of the recently developed
steady-state \textit{ab initio} laser theory (SALT)
\cite{TurStoGe2007,tureci_strong_2008,*TurStoRot2009,GeChoSto2010},
which provides the stationary solutions of the semiclassical
Maxwell-Bloch equations.  We can therefore make use of several parts
of the efficient methods developed for SALT in the optimization
problem at hand. In particular, we will employ the so-called threshold
constant flux (TCF) states $u_n$ defined by the non-Hermitian
eigenvalue problem \cite{GeChoSto2010},
\begin{equation}
\left\{ \nabla^2 + k^2 \,[ \varepsilon_c(\mathbf{x}) +
\eta_n(k) F(\mathbf{x})]
\right\} u_n(\mathbf{x},k) = 0\,.
\label{eq:gtcf}
\end{equation}
These states $u_n(\mathbf{x},k)$ and the corresponding eigenvalues
$\eta_n(k)$, which we parametrize over the wave number $k$ outside of
the disk, are self-orthogonal with respect to the pump profile
$F(\mathbf{x})$.  They furthermore have the important property that
each \textit{threshold laser mode} (TLM) $\bar{\Psi}_i$, which we are
interested in, is identical to one of these states,
$u_{\bar{n}}(\mathbf{x})$. (Note that, throughout the Letter, we will
use overbars to denote quantities at the lasing threshold.) The
corresponding required pump strength $\bar{D}_0^i$ and frequency
$\bar{k}_i$ at threshold are obtained from the threshold condition
$\eta_{\bar{n}}(\bar{k}_i)=\gamma(\bar{k}_i)\bar{D}_0^i$
\cite{liertzer_pump-induced_2012,GeChoSto2010}. We solve
\eqref{eq:gtcf} using a high-order finite element method which
efficiently copes with the finite-sized scatterers and the spatially
varying pump profile $F(\mathbf{x})$ \cite{*[{We employ the open
    source finite element mesher \emph{netgen} (}] [{) and the
    corresponding solver NGsolve available from
    \url{http://sf.net/projects/ngsolve} }] schoberl_netgen_1997}.

\begin{figure}
  \centering
  \includegraphics{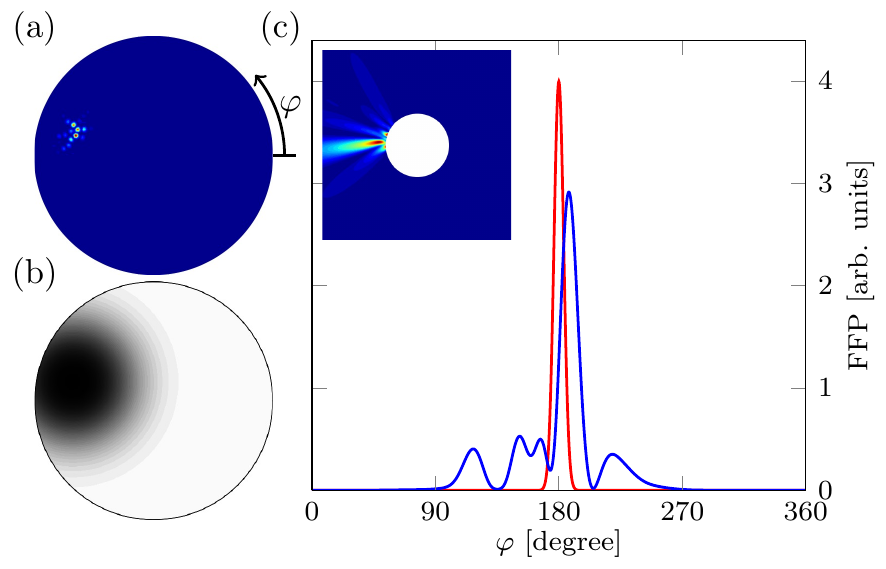}
  \caption{(color online) First threshold laser mode $\bar{\Psi}_{1}$
    at $\bar{k}_1=\myunitinv{29.82}$ of a random laser in the strongly
    scattering regime (see Fig.~\ref{fig:geom} with
    $n_{s}=3.0$). Shown are (a) the mode profile $|\fTLM{}(\xvec{})|$
    inside the random laser, (b) the Gaussian-shaped \pp{} that
    selects the mode (darker means more pump) and (c) the \ff{}
    profile (wiggly curve, blue) as well as the desired emission
    profile used as a reference (narrow peak, red). The inset in (c)
    depicts the near field directly outside the random laser disk and
    is complementary to (a). Note that the best directionality of a
    localized mode ($\mathcal{D} = 48.19\%$) matches the desired
    profile only approximately. The employed \gc{} has a center
    frequency at $k_{a}=\myunitinv{30.0}$ and a width of
    $\gT=\myunitinv{0.2}$. For better visibility the linear color
    functions used in the plots differ in absolute scale.}
  \label{fig:strong}
\end{figure}

The task that we address within the above framework is to shape the
emission pattern of the TLM which has the lowest (first) pump
threshold $\bar{D}_0^1$. Considering the scatterers of our disk-shaped
random medium $\varepsilon_c(\mathbf{x})$ as well as the gain curve
$\gamma(k)$ to be immobile (as in most experiments), we only tune the
applied pump profile $F(\mathbf{x})$ such that the laser emission at
threshold becomes unidirectional.  To quantify the unidirectionality
of a laser mode we evaluate its far-field profile FFP$(\varphi)$ for a
certain pump configuration, which is characterized by a set of complex
expansion coefficients $\bmnwb$ in a Bessel-function basis that
respects the circular symmetry of the disk boundary
\footnotetext[1]{See Supplemental Material at [] for the definition of
  the employed pump basis, the frequency variation of the first TLM in
  the course of the optimization procedure and for an efficient
  evaluation of the directionality gradient $\nabla \mathcal{D}$}
\footnotemark[1]. Next, we introduce a weight $\mathcal{D}$ which
measures the overlap of the far-field profile FFP$(\varphi)$ of this
mode with a desired profile $G(\varphi)$,
\begin{equation}
  \label{eq:measured2}
  \mathcal{D}(\bar{\Psi}_1) = \frac{\int_{0}^{2\pi} G(\varphi)\,
    {\rm FFP}[\varphi;\bar{\Psi}_1(\mathbf{x})]\,d\varphi}
  {\sqrt{\int_{0}^{2\pi}G^{2}\,d \varphi\,
      \int_{0}^{2\pi}{\rm FFP}^{2}\,d \varphi}}\,,
\end{equation}
where the denominator ensures that the weight is normalized such that
$\mathcal{D}\in[0,1]$. Choosing for the target function $G(\varphi)$ a
narrow Gaussian profile centered around $\varphi=180\degree{}$ [see
peak in \fig{strong}(c)], we can formulate the goal of unidirectional
emission in this direction as an optimization problem for the
coefficients $\bmnwb$ to be tuned such that $\mathcal{D}\to 1$.

Before going into the details of the optimization procedure, let us
first generally assess the applicability of our approach for a
dielectric featuring the two competing terms,
$\varepsilon=\varepsilon_c+\chi_g$, of the passive and the active
medium, respectively. Depending on which one of the two terms
dominates, the structure of the lasing modes will be determined either
by the passive medium or by the pump function. Correspondingly, our
approach based on controlling the pump profile will have to be
customized for each of these two limits.

In the regime of \textit{strong} disorder, where $\varepsilon_c$
dominates over $\chi_g$, the disorder itself will strongly trap modes
with, correspondingly, a high Q factor and a low lasing threshold. The
structure of these lasing modes is thus mostly determined by the
distribution of scatterers and not by the pump profile. Rather than
{\it optimizing} the latter, we can here just choose a pump profile
that simply {\it selects} the mode with the smallest deviation from
the desired unidirectional far-field pattern to be the first lasing
mode
\cite{andreasen_control_2010,leonetti_active_2013,vanneste_selective_2001}.

The situation is very different in the opposite limit of \textit{weak}
disorder, where $\chi_g$ dominates over $\varepsilon_c$. As no high-Q
modes exist from the outset, also mode selection does not
work. Rather, the pump profile has to be optimized such as to
\textit{shape} the modes in a way to be explored below.  Since the
case of strong disorder yields directional emission in a
straightforward way, it provides a natural benchmark to gauge the
quality of the emission patterns obtained for weakly disordered
samples. We will thus describe both the weak and the strong disorder
case, which we realize by a different contrast between the refractive
index of the rectangular scatterers $n_s$ in our sample and the
background index, $n_b=1$.

Let us first examine the limit of strong disorder, for which we choose
$n_s=3$. In this limit, the TCF states $u_n(\mathbf{x},k)$ resulting
from \eq{gtcf} with uniform pumping, $F(\mathbf{x})=$ const., are
strongly localized in space \cite{vanneste_selective_2001}. Among
these high-Q modes we also find many modes near the boundary of the
disorder region with, correspondingly, a very narrow emission profile.
To achieve the goal of directional emission, e.g., in the direction of
$\varphi=180\degree{}$, we just need to focus the pump beam on the
mode that emits most pronouncedly in this direction to make it lase
first \cite{vanneste_selective_2001}. We demonstrate this explicitly
with the example shown in \fig{strong}, where a directional lasing
mode is considered, which is the second TLM for the uniformly pumped
device.  Focusing a Gaussian shaped pump beam on the center of the
region where this mode is localized, we can rearrange the modes such
that this desired mode has the lowest lasing threshold of all. We have
explicitly tested this pump selection procedure for a number of modes
with different orientation of directional emission and found it to
work very well, provided that enough directional modes are available
and the gain curve is broad enough to be able to select them.

\begin{figure}
  \centering
  \includegraphics{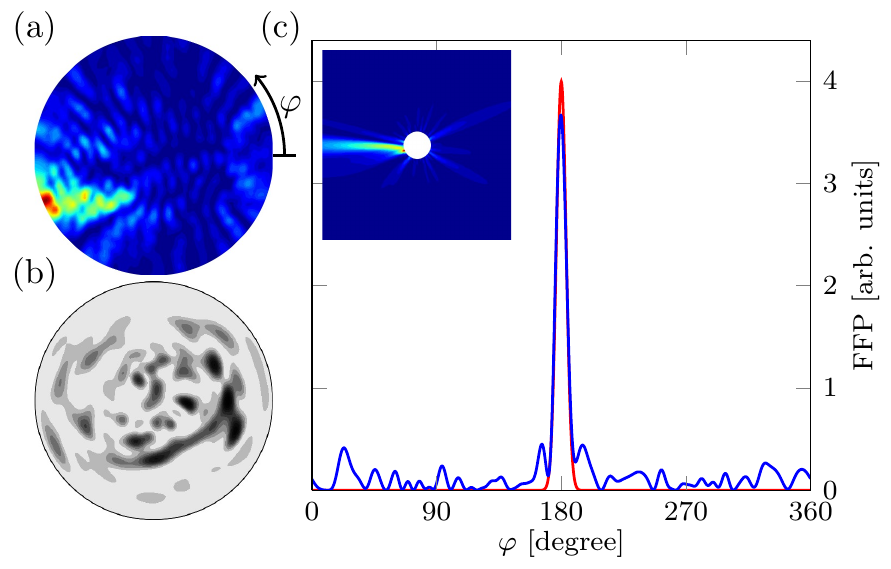}
  \includegraphics{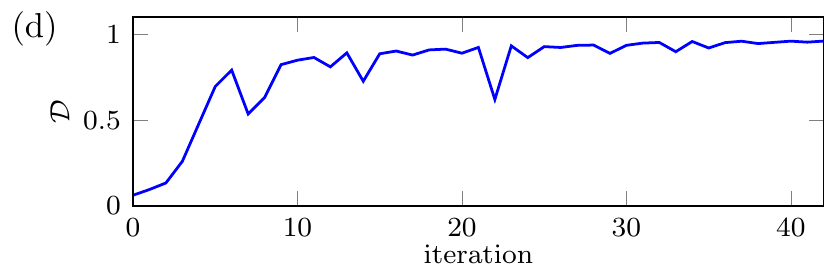}
  \caption{(Color online) First threshold laser mode $\bar{\Psi}_{1}$
    at $\bar{k}_{1}=\myunitinv{30.00}$ of a random laser in the
    \emph{weakly} scattering regime (see Fig.~\ref{fig:geom} with
    $n_{s}=1.2$).  Shown are (a) the optimized mode profile
    $|\fTLM{}(\xvec{})|$ inside the random laser, (b) the
    corresponding pump profile resulting from the optimization
    algorithm and (c) the \ff{} profile featuring a directionality of
    $\mathcal{D}=96.08\%$.  The inset in (c) shows the near field of
    the laser mode, just outside the pumped disk. Note the very good
    agreement which we find in the main panel of (c) between the
    optimized emission pattern (wiggly curve, blue) and the targeted
    Gaussian emission profile $G(\varphi)$ (narrow peak, red) centered
    at $\varphi=180\degree{}$ and with a width of $8.3\degree{}$.  In
    (d) the directionality is shown with respect to the number of
    iterations used in the optimization procedure. The parameters of
    the employed \gc{} are $k_{a}=\myunitinv{30.0}$ and
    $\gT=\myunitinv{0.01}$.}
  \label{fig:weak}
\end{figure}

In the weak disorder limit, for which we choose $n_s=1.2$, the lasing
modes are extended, overlapping and their shape strongly depends on
the pump profile $F(\mathbf{x})$
\cite{tureci_strong_2008,*TurStoRot2009}. Simple mode selection is
thus ruled out and a more sophisticated
optimization is required instead.  We start with an adjustable pump profile
characterized by $153$ complex expansion coefficients $\bmnwb$
\footnotemark[1]. An optimization procedure is now employed to
determine these parameters such that the far-field emission profile
FFP$(\varphi)$ follows the desired narrow shape $G(\varphi)$ or,
equivalently, $\mathcal{D}\to 1$ in Eq.~(\ref{eq:measured2}). Note
that our algorithm always optimizes the first TLM (which is determined
for each iteration step anew) such that our laser stays single mode
during the optimization procedure. Note also that it is clearly
unfeasible to find the \textit{unique} global maximum for
$\mathcal{D}(\bar{\Psi}_1)$ in the extremely large parameter space of
the above complex coefficients.  Instead, we start with random initial
guesses for the pump profile coefficients $\bmnwb$ which we further
refine locally by a gradient-based optimization \cite{*[{We employ the
    \emph{nlopt} toolkit using the Multi-Level Single-Linkage
    algorithm as the global optimization routine where a
    low-discrepancy sequence is used for the generation of random
    initial guesses. The method of moving asymptotes is used for the
    gradient-based local optimization. }] [{}]
  johnson_nlopt_2012,*svanberg_class_2002,*kan_stochastic_1987,*kan_stochastic_1987-1,*kucherenko_application_2005}.
For the local optimization procedure one needs to evaluate the
gradient of the directionality $\nabla\mathcal{D}$ with respect to the
expansion coefficients $\bmnwb$. Because of the large number of involved
parameters, it is crucial to devise a very fast and sufficiently
accurate scheme to evaluate the gradient, which task can be solved
with the help of the TCF state basis and its self-orthogonality
property \footnotemark[1].\begin{figure}
  \centering
  \vspace{2mm}
  \includegraphics{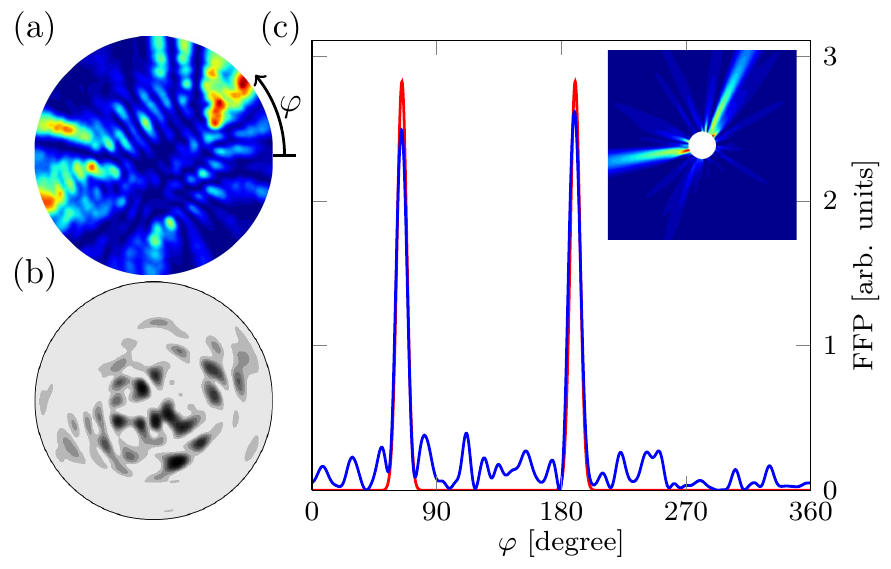}
  \caption{(Color online) Results from the optimization of the random
    laser for emitting into two prescribed directions ($65\degree$ and
    $190\degree$) simultaneously (the same disorder is used as in
    \fig{weak}). Shown are (a) the optimized mode profile
    $|\fTLM{}(\xvec{})|$ inside the random laser, (b) the
    corresponding pump profile resulting from the optimization
    algorithm and (c) the \ff{} profile featuring a convergence degree
    of $\mathcal{D}=96.12\%$.  The inset in (c) shows the near field
    of the laser mode, just outside the pumped disk. Note the very
    good agreement which we find in the main panel of (c) between the
    optimized emission pattern (wiggly curve, blue) and the targeted
    double-peak emission profile $G(\varphi)$ (narrow peaks, red).}
  \label{fig:twodirs}
\end{figure}

In \fig{weak}(d) we display the progression of the directionality
$\mathcal{D}$ of the first TLM during the optimization procedure which
leads to a rather quick convergence (in 42 iterative steps) from an
initially random starting point to a nearly optimal mode with the
desired directionality. The directionality does not increase
monotonically as both the gradient of the directionality is only
approximate and the first TLM can switch between modes
\footnotemark[1]. Note also that typically several random starting
conditions for the iteration are necessary due to the existence of
local maxima of $\mathcal{D}(\bmnwb)$ in the multidimensional
parameter space of $\bmnwb$ where the local optimization may get
stuck. More sophisticated methods to overcome local maxima could
certainly be applied here. Since, however, we already do reach
excellent results with $\mathcal{D}>96\%$ with typically not more than
altogether 20 initial guesses, such methods are not compulsory.  The
mode which matches the emission pattern of the target function
$G(\varphi)$ best and the corresponding pump profile are shown in
\fig{weak}. This encouraging result is directly reflected by the shape
of the corresponding lasing mode shown in \fig{weak}(a) and
\ref{fig:weak}(c) which displays the desired narrow, beamlike shape
outside of the disk boundary. We emphasize, in particular, that the
directionality can here be much better controlled than in strongly
disordered media. This is best observed by comparing
Figs.~\ref{fig:strong} and \ref{fig:weak}: Although the mode which
agrees best with the desired emission profile $G(\varphi)$ is selected
in the strongly scattering medium (\fig{strong}), a clear deviation
from the desired target function $G(\varphi)$ can be observed. In
fact, for disorder configurations different than the one in
\fig{strong} we observed much larger deviations. For weakly scattering
media, on the contrary, the applied pump profile is optimized for each
specific distribution of scatterers such that a directionality of
$D\gtrsim 95\%$ is generally guaranteed.

Our optimization procedure can be pushed even further in the sense
that the reference emission pattern $G(\varphi)$ is not limited to a
function with a single narrow peak, but instead can be chosen almost
arbitrarily. As an example we have also optimized the pump profile 
such that the random laser emits into two
directions simultaneously; see \fig{twodirs}.
This result shows the versatility of our approach and manifests that
the emission of a weakly scattering random laser can be arbitrarily
shaped using a carefully optimized pump profile. Combining this
technique together with the frequency selection mechanism presented in
\cite{bachelard_taming_2012,bachelard_turning_2013} suggests that
random lasers can be controlled on an unprecedented scale.

We emphasize that the pump profiles which give rise to these uni- or
multidirectional lasing states are neither very focused (as in
\cite{vanneste_selective_2001,leonetti_active_2013}), nor strip-shaped
(as in \cite{wu_random_2006,ge_steady-state_2010-1}), nor in any other
way evidently self-explanatory.  Instead, the obtained profiles are
structured on a scale comparable to the lasing wavelength -- a
situation that is routinely created in laboratory experiments
\cite{vellekoop_focusing_2007,KatSmaBro2011,AulGjoJoh2011,popoff_image_2010,*popoff_measuring_2010,bachelard_turning_2013}.
The pump optimization procedure presented above can, in principle, be
mapped to a corresponding experiment using, e.g., a spatial light
modulator for shaping the pump beam (see Fig.~\ref{fig:geom}). In this
case the optimization can be organized through a feedback loop between
the measured emission and the applied pump profile, as implemented
very recently in a similar setting \cite{bachelard_turning_2013}.
Realizing such ideas in the experiment would not only be interesting
for fundamental research on light profile patterning in a disordered
medium, but would constitute a major step forward in bringing random
lasers closer to practical applications where specific emission
profiles are required.

\begin{acknowledgments}
  The authors would like to thank N. Bachelard, S. Gigan and P. Sebbah
  for carefully reading our manuscript and the following colleagues
  for very fruitful discussions: S.~Esterhazy, D.~Krimer,
  J.~M.~Melenk, and J.~Sch\"oberl.  Financial support by the Vienna
  Science and Technology Fund (WWTF) through Project No. MA09-030, by
  the European Research Council (ERC) (project ODYQUENT) and by the
  Austrian Science Fund (FWF) through Projects No.~F25-14 (SFB IR-ON),
  No.~F49-P10 (SFB NextLite) is gratefully acknowledged. We are also
  indebted to the administration of the Vienna Scientific Cluster
  (VSC) for generously granting us access to computational resources.
\end{acknowledgments}

\end{document}


\title{Supplemental material for pump-controlled directional light
  emission from random lasers}

\author{Thomas Hisch}
\email{thomas.hisch@tuwien.ac.at}
\affiliation{Institute for Theoretical Physics, Vienna University of
             Technology, Wiedner Hauptstra\ss e 8--10/136, 1040 Vienna, Austria, EU}

\author{Matthias Liertzer}
\affiliation{Institute for Theoretical Physics, Vienna University of
             Technology, Wiedner Hauptstra\ss e 8--10/136, 1040 Vienna, Austria, EU}

\author{Dionyz Pogany}
\affiliation{Institute for Solid State Electronics, Vienna University of Technology, Floragasse 7, 1040 Vienna, Austria, EU}

\author{Florian Mintert}
\affiliation{Freiburg Institute for Advanced Studies, Albert--Ludwigs University of Freiburg, Albertstra\ss e 19, 79104 Freiburg, Germany, EU}

\author{Stefan Rotter}
\email{stefan.rotter@tuwien.ac.at}
\affiliation{Institute for Theoretical Physics, Vienna University of
              Technology, Wiedner Hauptstra\ss e 8--10/136, 1040 Vienna, Austria, EU}

\date{\today}

\maketitle

\section*{Pump profile basis}
To first specify and then optimize the pump profile for directional
emission we expand the two-dimensional pump profile $F(\mathbf{x})$ on
the disk-shaped random laser of radius $R$ using a Bessel function
basis of the form $J_m(j_{m,n}|\xvec{}|/R) \exp[-i m\varphi]$. Here
the $j_{m,n}$ are the Bessel function roots, $\varphi$ the polar
angle, and $R$ the radius of the disk within which the scatterers are
randomly distributed.  This basis has the suitable property of
vanishing at the disk boundary outside of which no pump is
applied. Inside the disk we expand the real and strictly positive pump
function $F(\xvec{})$ as follows,
\begin{equation}
  F(\xvec{};\bmn) \!=\! \left|\sum_{m,n} \bmn
    J_{m}(j_{m,n}\frac{|\xvec{}|}{R}) \exp[-im\varphi]\right|^{2}\!,
  \label{eq:bbasis_orig}
\end{equation}
where $\bmn$ are complex expansion coefficients. In our numerical
calculations we restrict the summation in Eq.~(\ref{eq:bbasis_orig})
to the limits $m\in[-m_{\text{max}},m_{\text{max}}]$ and
$n\in[0,n_{\text{max}}]$, corresponding to a finite resolution of the
pump beam in radial and azimuthal direction, respectively. The choice
which we make for the
$(2m_{\text{max}}\!+\!1)\times(n_{\text{max}}\!+\!1)$ complex
coefficients $\bmn$ will determine the threshold laser mode
$\bar{\Psi}_1(\xvec{})$. In the calculations presented in the main
part of the article (see Fig.~3), we choose $n_{\text{max}}=8$ and
$m_{\text{max}}=8$. Note that, although not explicitly
  shown here, one can reduce the complex parameters $\bmn$ to purely
  real ones since only the absolute square of the superposition of
  states is considered.

\section*{Emission frequency behavior}

\begin{figure}[b]
\includegraphics{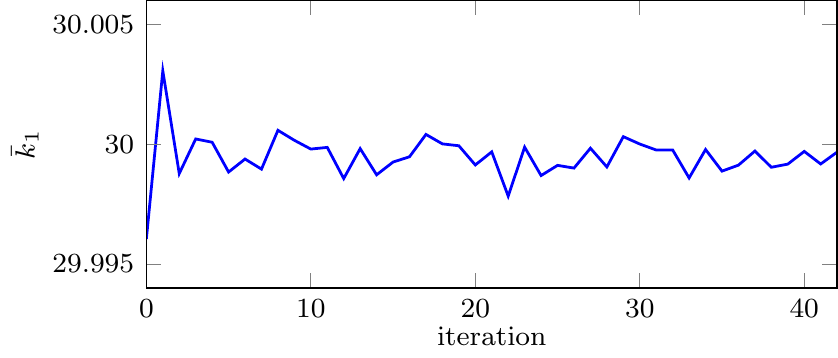}
\caption{Variation of the emission frequency $\bar{k}_1$ of the first
  threshold laser mode in the course of the optimization procedure
  (the same parameters were used here as for Fig.~3d in the main
  text).}
\label{fig:freq_behavior}
\end{figure}

In the course of optimizing the directionality of the first TLM the
changes in the pump profile also lead to slight variations of the
laser frequency around the peak gain frequency of
\SI{30}{${\mu}m^{-1}$}. For the specific case displayed in Fig.~3d of
the main text this frequency variation takes on the form shown in
Fig.~\ref{fig:freq_behavior}. Note that for the first few iteration
steps this variation is larger since the pump profile is modified more
strongly here as compared to later steps of the optimization
procedure.

\section*{Derivation of the gradient of the directionality}In order to efficiently calculate the gradient of the directionality
$\mathcal{D}$ with respect to the complex coefficients $\bmnwb$ we
approximate this gradient with the help of the TCF states which are
defined as follows
\begin{equation}
  \left\{ \nabla^2 + \bar{k}_1^2 \,[ \varepsilon_c(\mathbf{x}) +
    \eta_n(\bar{k}_1) F(\mathbf{x},\bar{k}_1)]
  \right\} u_n(\mathbf{x}) = 0\,,
  \label{eq:gtcf}
\end{equation}
and which satisfy the self-orthogonality relation
\begin{equation}
  \int_{r<R} u_i(\mathbf{x}) F(\mathbf{x}) u_j(\mathbf{x})
  \mathrm{d}\mathbf{x} = \delta_{i,j}.
  \label{eq:selfortho}
\end{equation}
The basis itself is calculated at the frequency $\bar{k}_1$ of the
first threshold laser mode $\bar{\Psi}_1$, for which we optimize the
angular emission pattern to match the targeted emission pattern
$G(\varphi)$. In order to reduce the terminology in the following we
rewrite, without loss of generality, the set of complex coefficients
$\bmnwb$ as a set of twice as many real coefficients $\alpha_i$, where
the imaginary part of the coefficients $\bmnwb$ is moved into the
corresponding basis function $f_i(\mathbf{x})$, such that,
\begin{eqnarray}
  F(\xvec{};\bmn) &=& \left|\sum_{m,n} \bmn
    J_{m}(j_{m,n}\frac{|\xvec{}|}{R})
    \exp[-im\varphi]\right|^{2}\!\\
  &=:& \left|\sum_i \alpha_i f_i(\mathbf{x})\right|^2.
  \label{eq:bbasis}
\end{eqnarray}

We will now derive the gradient of the directionality $\mathcal{D}$
with respect to these coefficients $\alpha_i$. The first part of this
derivation follows directy from differentiation of $\mathcal{D}$ with
respect to $\alpha_j$,
\begin{eqnarray}
  \frac{\partial}{\partial \alpha_j} \mathcal{D}[\{\alpha_i\}] &=&
  \partial_{\alpha_j} \int \tilde{G}(\varphi)
  \frac{\flux(\{\alpha_i\},\varphi)}{\sqrt{\int
      \flux^2(\{\alpha_i\},\varphi') \mathrm{d}\varphi'}}
  \mathrm{d}\varphi  \\
  &=& \int \tilde{G}(\varphi) \biggl[ \frac{\partial_{\alpha_j}\flux(\{\alpha_i\},\varphi)}{\sqrt{\int
      \flux^2(\{\alpha_i\},\varphi')
      \mathrm{d}\varphi'}} -\\
  &\,&\qquad\,\,-\,\,\,\frac{\flux(\{\alpha_i\},\varphi)
    \int{\flux(\{\alpha_i\},\varphi')\partial_{\alpha_j}\flux(\{\alpha_i\},\varphi')\mathrm{d}\varphi'}}{\left(\int
      \flux^2(\{\alpha_i\},\varphi') \mathrm{d}\varphi'\right)^{\frac{3}{2}}
  }\biggr] \mathrm{d}\varphi,\label{eq:gradientderiv}
\end{eqnarray}
where $\tilde{G}(\varphi) = G(\varphi)/\sqrt{\int
  G(\varphi')^2\mathrm{d}\varphi'}$ is the normalized reference
emission profile and $\flux$ is the angular far-field emission profile
of the first TLM. The latter is determined as the real part of the
complex Poynting vector at $r\to\infty$ which can be simplified to
\begin{equation}
  \label{eq:poynting}
  \flux(\varphi):=
  \lim_{r\to\infty}\frac{r}{2\bar{k}_1}\text{Im}[\bar{\Psi}_1^*(r,\varphi)\partial_r\bar{\Psi}_1(r,\varphi)]\,.
\end{equation}
The partial derivative of $\flux$ with respect to the coefficients
$\alpha_j$ is given by
\begin{equation}
\partial_{\alpha_j}\flux(\{\alpha_i\},\varphi)=  \lim_{r\to\infty}
\frac{r}{2\bar{k}_1}\imag\left[(\partial_{\alpha_j}\bar{\Psi}_{1}^{*})\partial_{r}\bar{\Psi}_{1}
  +
  \bar{\Psi}_{1}^{*}\partial_{r}(\partial_{\alpha_j}\bar{\Psi}_{1})\right]. \label{eq:fluxderiv}
\end{equation}

For reasons of improved efficiency we express the gradient of the TLM
$\partial_{\alpha_j}\bar{\Psi}_{1}$ in terms of the TCF states
$u_n$. Note, that the first TLM $\bar{\Psi}_1$ is equivalent to a TCF
state $u_{\bar{n}}$ evaluated at the threshold laser frequency
$\bar{k}_1$, i.e., $\bar{\Psi}_1({\bf x})=u_{\bar{n}}({\bf
  x},\bar{k}_1)$. With the help of this equivalence the gradient of
$\bar{\Psi}_1$ is formally defined as
\begin{equation}
  \partial_{\alpha_j}u_{\bar{n}} = \lim_{h\to0}\frac{u_{\bar{n}}(\{\alpha_i + h
    \delta_{i,j}\}) - u_{\bar{n}}(\{\alpha_i\})}{h}.
\end{equation}
We can now proceed and use perturbation theory in order to derive an
expression for $\partial_{\alpha_j}u_{\bar{n}}$ explicitly. With the
operator $\hat{L} = -\nabla^2 - \bar{k}_1^2\varepsilon_c$ we can rewrite the
original eigenvalue problem Eq.~\eqref{eq:gtcf} as
\begin{equation}
  \hat{L} u_n = \eta_n \bar{k}_1^2 F(\{\alpha_i\}) u_n \label{eq:tcf2}
\end{equation}
and the ``perturbed'' eigenvalue problem as
\begin{equation}
  \hat{L} \tilde{u}_n = \tilde{\eta}_n \bar{k}_1^2 F(\{\alpha_i\ +
  h\delta_{i,j}\}) \tilde{u}_n, \label{eq:perttcf}
\end{equation}
where the perturbed state $\tilde{u}_{n} = u_{n}(\{\alpha_i + h
\delta_{i,j}\})$ and the perturbed eigenvalue $\tilde{\eta}_{n} =
\eta_{n}(\{\alpha_i + h \delta_{i,j}\})$. The perturbed pump profile can
be approximated by
\begin{eqnarray}
F(\{\alpha_i+h\delta_{i,j}\}) &=& \left|\sum_i\alpha_i f_i(x) + hf_j(x)\right|^2 \\ 
  &=& F(\{\alpha_i\}) + h \underbrace{\sum_{i} 2 \alpha_{i}
    \real({f_{i}f_{j}^{*}})}_{F'(\{\alpha_i\})} + \mathcal{O}(h^2). \label{eq:pertpump}
\end{eqnarray}

In a next step we insert this approximation, together with the
following ansatz for the perturbed eigenstate,
\begin{equation}
\tilde{u}_{\bar{n}} \approx u_{\bar{n}} + h\sum_{\substack{i=1\\ i \neq \bar{n}}}^{N} c_{i,\bar{n}} u_i, \label{eq:pertansatz}
\end{equation}
and for the eigenvalue, $\tilde{\eta}_n = \eta_n + h \eta_n'$, into
Eq.~\eqref{eq:perttcf} ($N$ denotes here the size of the TCF
basis). By discarding terms of $\mathcal{O}(h^2)$ and making use of
the self-orthogonality of the TCF-states [Eq.~\eqref{eq:selfortho}] we
obtain the expansion coefficients of the perturbed TCF state,
\begin{equation}
c_{i,\bar{n}} = \frac{\eta_{\bar{n}}}{\eta_i-\eta_{\bar{n}}} \int_{r<R} u_i(\mathbf{x}) F'(\{\alpha_i\})
u_{\bar{n}}(\mathbf{x}) \mathrm{d}\mathbf{x},
\end{equation}
where $R$ is the radius of our random laser disk. This expression can then be
inserted back into Eq.~\eqref{eq:pertansatz}, which yields
\begin{equation}
  \partial_{\alpha_j}u_{\bar{n}} = \sum_{\substack{i=1\\ i \neq \bar{n}}}^{N}c_{i,\bar{n}}u_i.
\end{equation}
With this expression and Eq.~\eqref{eq:fluxderiv}, the desired approximate gradient of the
directionality measure $\mathcal{D}$ with respect to the coefficients
$\alpha_i$, Eq.~(\ref{eq:gradientderiv}), is now easily evaluated.